\documentclass[aps,prd,groupaddress,tighten,nofootinbib,twocolumn]{revtex4}
\usepackage{amssymb}
\usepackage{mathrsfs}
\usepackage{mathbbold}
\usepackage{graphicx}
\usepackage{amsmath}
\usepackage{epsfig}
\usepackage{color}
\def\slc#1{\setbox0=\hbox{$#1$}           
    \dimen0=\wd0                                 
    \setbox1=\hbox{/} \dimen1=\wd1               
    \ifdim\dimen0>\dimen1                        
       \rlap{\hbox to \dimen0{\hfil/\hfil}}      
       #1                                        
    \else                                        
       \rlap{\hbox to \dimen1{\hfil$#1$\hfil}}   
       /                                         
    \fi}

\begin{document}


\title{Non-Standard Neutrino Interactions from a Triplet Seesaw Model}

\author{Michal Malinsk\'y}
\email{malinsky@kth.se}

\author{Tommy Ohlsson}
\email{tommy@theophys.kth.se}

\author{He Zhang}
\email{zhanghe@kth.se}

\affiliation{Department of Theoretical Physics, School of
Engineering Sciences, Royal Institute of Technology (KTH) --
AlbaNova University Center, Roslagstullsbacken 21, 106 91 Stockholm,
Sweden}


\begin{abstract}
We investigate non-standard neutrino interactions (NSIs) in the
triplet seesaw model featuring non-trivial correlations between NSI
parameters and neutrino masses and mixing parameters. We show that
sizable NSIs can be generated as a consequence of a
nearly degenerate neutrino mass spectrum. Thus, these NSIs could lead to quite significant signals of lepton flavor violating
decays such as $\mu^- \rightarrow e^- {\nu}_e \overline{\nu_\mu}$ and $\mu^+ \rightarrow e^+ \overline{\nu_e} \nu_\mu$ at a future
neutrino factory, effects adding to the uncertainty in determination of the
Earth matter density profile, as well as characteristic patterns of the doubly charged Higgs decays observable at the Large Hadron Collider.
\end{abstract}

\maketitle

\section{Introduction}

Recently, a large number of studies has been dedicated to the
phenomenology of neutrino oscillations and non-standard neutrino interactions
(NSIs) emerging from the effective Lagrangian of the form
\begin{eqnarray}\label{eq:L-NSI}
{\cal L}_{\rm NSI} = -2\sqrt{2}{G_F}
\varepsilon^{ff'C}_{\alpha\beta} \left(
\overline{\nu_\alpha}\gamma^\mu P_{L} \nu_\beta \right) \left(
\overline{f}\gamma_\mu P_{C} f' \right)\,,
\end{eqnarray}
where $f$ and $f'$ denote charged lepton or quark fields, $G_F$ is
the Fermi coupling constant, and $P_{L}$ and $P_{C}$ (with $C=L, R$)
stand for different chiral projectors. It is generally expected that
the effective operator 
comes out of an
underlying theory respecting (or even encompassing) the Standard
Model (SM) gauge symmetry upon integrating out a certain set of
``heavy'' degrees of freedom.

A vast majority of the previous works  was concerning the
matter-induced NSIs or NSIs at source and/or detector at the level
of the effective operator (\ref{eq:L-NSI}). Here, instead, we focus
on a particular extension of the SM featuring an extra
$SU(2)_{L}$-triplet Higgs
\cite{Schechter:1980gr,Lazarides:1980nt,Mohapatra:1980yp}, which
provides a very popular and simple scheme for accommodating Majorana
masses of neutrinos within a renormalizable framework, and indeed,
can induce significant NSI effects in a future neutrino factory.

This work is organized as follows. In Sec.~\ref{Sec:1}, we present
the relations between the effective NSI parameters and the triplet
Yukawa couplings. Next, Sec.~\ref{Sec:2} is devoted to a thorough
discussion of the experimental constraints stemming namely from
exotic charged lepton decays, giving rise to upper bounds for the
NSI parameters. Then, in Sec.~\ref{Sec:3}, we provide a simple
estimate of the possible effects at a neutrino factory and at the Large
Hadron Collider (LHC). Finally, in Sec.~\ref{Sec:4}, we summarize
our results and conclude.

\begin{figure}[t]
\begin{center}
\ensuremath{\vcenter{\hbox{\includegraphics[width=8cm]{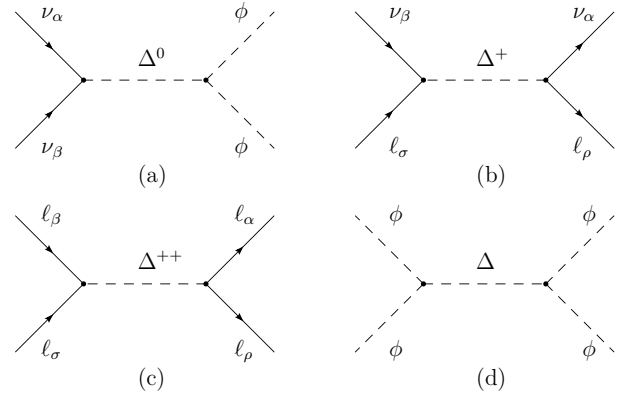}}}}
\caption{Tree
level Feynman diagrams with the exchange of heavy triplet fields.
The corresponding diagrams are responsible for (a) a light neutrino
Majorana mass term, (b) non-standard neutrino interactions, (c) NSIs
of four charged leptons, and (d) self-coupling of the SM Higgs
doublet.} \label{fig:fig1} \vspace{-0.5 cm}
\end{center}
\end{figure}
\section{Light neutrino mass matrix and non-standard interactions}
\label{Sec:1}

We consider the simple extension of the SM with an extra Higgs
triplet \cite{Schechter:1980gr,Lazarides:1980nt,Mohapatra:1980yp}.
Gauge invariance allows one to add the couplings between the Higgs
triplet and two lepton doublets as well as a coupling of the scalar
triplet to the SM Higgs doublet \cite{Chao:2006ye}
\begin{eqnarray}\label{eq:L-Y}
{\cal L}_{\Delta} =   Y_{\alpha\beta} L^{T\alpha}_L C\,{\rm
i}\sigma_2 \Delta L^\beta_L + \lambda_\phi  \phi^T {\rm i} \sigma_2
\Delta^\dagger \phi + {\rm H.c.}\ ,
\end{eqnarray}
where $\Delta$ is a $2\times 2$ representation of the Higgs triplet
field
\begin{eqnarray}\label{eq:delta}
\Delta=\left (\begin{matrix}\displaystyle {\Delta^+/ \sqrt{2}} &
\displaystyle \Delta^{++}_{}\cr \displaystyle \Delta^0_{}&
\displaystyle -{\Delta^+ / \sqrt{2}} \end{matrix}\right) \ ,
\end{eqnarray}
and $Y$ is a $3\times3$ symmetric matrix in flavor space.
Written in component fields, the Yukawa couplings in
Eq.~\eqref{eq:L-Y} receive the form
\begin{eqnarray}\label{eq:Yukawa}
{\cal L}_{ Y } & = & Y_{\alpha\beta} \bigg[ \Delta^0 \nu_\alpha P_L
\nu_\beta - {1\over \sqrt{2}} \Delta^+ \left(
\overline{\ell^c_\alpha} P_L \nu_\beta + \overline{\nu^c_\alpha} P_L
\ell_\beta \right) \nonumber \\ &-& \Delta^{++}
\overline{\ell^c_\alpha} P_L \ell_\beta \bigg] + {\rm H.c.}
\end{eqnarray}
Integrating out the heavy triplet field at tree level (as
illustrated in Fig.~\ref{fig:fig1}), one can obtain effective
dimension five and six operators, which are responsible for light
neutrino masses and NSIs, respectively,
\cite{Chao:2006ye,Abada:2007ux,Gavela:2008ra}
\begin{eqnarray}\label{eq:mass}
{\cal L}_{\nu}^{m} & = &  \frac{ Y_{\alpha\beta} \lambda_\phi
v^2}{m^2_\Delta}\left(\nu^c_{L\alpha}  \nu_{L\beta} \right) =   -
{1\over 2}  (m_\nu)_{\alpha\beta} \nu^c_{L\alpha}
 \nu_{L\beta}  \ ,\ \ \  \\
\label{eq:NSI} {\cal L}_{\rm NSI} & = &  \frac{ Y_{\sigma\beta}
Y^\dagger_{\alpha\rho}}{m^2_{\Delta}}\left( \overline{\nu_\alpha}
\gamma_\mu P_L \nu_\beta \right)  \left( \overline{\ell_\rho}
\gamma^\mu P_L \ell_\sigma \right) \ , \\
 \label{eq:4l} {\cal
L}_{4\ell} & = & \frac{ Y_{\sigma\beta}
Y^\dagger_{\alpha\rho}}{m^2_{\Delta}}\left( \overline{\ell_\alpha}
\gamma_\mu P_L \ell_\beta \right)  \left( \overline{\ell_\rho}
\gamma^\mu P_L \ell_\sigma \right)  \ ,
\end{eqnarray}
where $v\simeq 174~{\rm GeV}$ is the vacuum expectation value of the
SM Higgs field. For the NSIs (\ref{eq:NSI}) to be potentially
sizable, we must require the triplet to be rather light, typically at
the TeV scale. Notice that  in such a case the dimensionful
parameter $\lambda_{\phi}$ associated with the trilinear Higgs
coupling must be small enough to keep the absolute neutrino mass
scale proportional to $\lambda_{\phi}v^{2}/m_{\Delta}^{2}$ in the
sub-eV range. However, this could be  natural, since the symmetry is
enhanced in the zero $\lambda_{\phi}$ limit \cite{tHooft:1979bh}.

Comparing Eqs.~\eqref{eq:mass} and \eqref{eq:NSI} with
Eq.~\eqref{eq:L-NSI}, we can establish relations between the light
neutrino mass matrix and NSI parameters in Eq. (\ref{eq:L-NSI}) as
(for left lepton chirality)
\begin{eqnarray}\label{eq:epsilon}
\varepsilon^{\rho\sigma}_{\alpha\beta} & = & -\frac{m^2_\Delta}{8
\sqrt{2}G_F v^4 \lambda^2_\phi} \left(m_\nu\right)_{\sigma\beta}
\left(m^\dagger_\nu\right)_{\alpha \rho} \ .
\end{eqnarray}
Equation~\eqref{eq:epsilon} clearly shows the correlations between
the NSI effects and standard oscillation parameters in the triplet
seesaw model. According to Eq.~\eqref{eq:epsilon}, it is easy to see
that the NSI parameters are not independent, and in fact, they are
strongly tied to the structure of the light neutrino mass matrix
$m_{\nu}$. However, not all the $\varepsilon$'s are physically
interesting parameters. For the propagation process in long baseline
experiments, neutrinos encounter Earth matter effects and only
electron type of NSIs $\varepsilon^m_{\alpha\beta} \equiv
\varepsilon^{ee}_{\alpha\beta}$ contributes to the matter potential.
In addition, Eq.~\eqref{eq:epsilon} affects neutrino production at
neutrino sources, especially for a neutrino factory. More generally,
both the processes $\mu^{-}\to e^{-}\nu_{\mu}\overline{\nu_{\beta}}$
(corresponding to the NSI parameter $\varepsilon^{e\mu}_{\mu\beta}$)
and $\mu^{-}\to e^{-}\nu_{\alpha}\overline{\nu_{\beta}}$ (corresponding to
$\varepsilon^{e\mu}_{\alpha\beta}$ with $\alpha \neq \mu$) may
occur, and their contributions have to be added to the SM transition
rate either coherently or incoherently
, depending on the specific situation. In the
following section, we will discuss the current experimental
constraints on these NSI parameters in detail.

\section{Constraints on NSI parameters}\label{Sec:2}

We now summarize the current experimental constraints on NSI
parameters defined in Eq.~\eqref{eq:epsilon}. The most stringent
experimental bounds come from the lepton family number violating
(LFV) decays $\mu \rightarrow 3e$ and $\tau \rightarrow 3 \ell$,
which are tree-level processes induced by ${\cal L}_{4\ell}$ in
Eq.~\eqref{eq:4l}. In terms of NSI parameters, the corresponding
decay widths are given by \cite{Abada:2007ux}
\begin{eqnarray}\label{eq:width}
\Gamma (\mu^- \rightarrow e^- e^+ e^-) & = & \frac{m^5_\mu}{24 \pi^3}
G^2_F \left| \varepsilon^{e\mu}_{ee} \right|^2 \ , \\
\Gamma (\tau^- \rightarrow \ell^-_\alpha \ell^+_\beta  \ell^-_\alpha)
& = & \frac{m^5_\tau}{24 \pi^3}
G^2_F \left| \varepsilon^{\alpha \tau}_{\alpha \beta} \right|^2 \ , \\
\Gamma (\tau^- \rightarrow \ell^-_\alpha \ell^+_\beta \ell^-_\rho)
& = & \frac{m^5_\tau}{12 \pi^3} G^2_F \left| \varepsilon^{\rho
\tau}_{\alpha\beta} \right|^2 \ .
\end{eqnarray}
Another type of stringent constraints on the current model comes
from the rare radiative lepton decays $\ell_\alpha \rightarrow
\ell_\beta \gamma$, although these processes emerge at one-loop
level. In neglecting the light charged lepton masses, we have
\cite{Bilenky:1987ty}
\begin{eqnarray}\label{eq:width2}
\frac{ \Gamma (\ell_\sigma^{-} \rightarrow \ell^{-}_\rho \gamma)}{\Gamma
(\ell^{-}_\sigma\rightarrow \ell^{-}_\rho  \nu_\sigma
\overline{\nu_\rho})} = \frac{\alpha}{6\pi} \frac{25}{16}\left|
\sum_\alpha \varepsilon^{\rho\sigma}_{\alpha\alpha} \right|^2 \ .
\end{eqnarray}

As for the diagonal parts of $Y$, strong bounds come from Bhabha
scattering and muonium to antimuonium conversion, which lead to
$|Y_{ee} Y_{\mu\mu}^{*}|<0.1 \times (m_\Delta/1~{\rm TeV})^2 $
\cite{Willmann:1998gd}. There are also constraints from the
universality of weak interactions, precise experimental measurements
of the $W$ boson mass and the $\rho$ parameter, which are relatively
weak compared to the bounds discussed above \cite{Antusch:2008tz}
and will not be elaborated on in our calculations.

The constraints are summarized in Table~\ref{tab:constraints}.
\begin{table}
\begin{center}
\begin{tabular}{|c|c|c|}
\hline
Decay & Constraint on & Bound
\\
\hline
$\mu^- \rightarrow e^- e^+ e^- $  &  $| \varepsilon^{e \mu}_{ee} |$ & $ 3.5 \times 10^{-7}$  \\
\hline $\tau^-
\rightarrow e^- e^+  e^- $  & $| \varepsilon^{e \tau}_{ee} |$  & $ 1.6 \times 10^{-4}$  \\
\hline
$\tau^- \rightarrow  \mu^- \mu^+ \mu^-$  & $| \varepsilon^{\mu \tau}_{\mu\mu} |$  & $ 1.5 \times 10^{-4}$  \\
\hline
$\tau^- \rightarrow e^- \mu^+  e^-$  & $| \varepsilon^{e \tau}_{e\mu} |$  & $1.2 \times 10^{-4}$  \\
\hline
$\tau^- \rightarrow \mu^- e^+  \mu^-$  & $| \varepsilon^{\mu \tau}_{\mu e } |$  &  $ 1.3 \times 10^{-4}$  \\
\hline
$\tau^- \rightarrow e^- \mu^+ \mu^-  $  & $| \varepsilon^{e \tau}_{\mu\mu} |$  &  $ 1.2 \times 10^{-4}$ \\
\hline
$\tau^- \rightarrow e^- e^+ \mu^-$  & $| \varepsilon^{e \tau}_{\mu e} |$  &  $ 9.9 \times 10^{-5}$ \\
\hline
$\mu^- \rightarrow e^- \gamma  $  &  $| \sum_{\alpha} \varepsilon^{e \mu}_{\alpha\alpha} |$ &  $ 1.4 \times 10^{-4}$ \\
\hline
$\tau^- \rightarrow e^- \gamma  $  & $| \sum_{\alpha} \varepsilon^{e \tau}_{\alpha\alpha} |$  &  $3.2 \times 10^{-2}$ \\
\hline
$\tau^- \rightarrow \mu^- \gamma  $  & $| \sum_{\alpha} \varepsilon^{\mu \tau}_{\alpha\alpha} |$  &  $2.5 \times 10^{-2}$ \\
\hline
$\mu^+ e^- \rightarrow \mu^- e^+$  &  $| \varepsilon^{\mu e}_{\mu e} |$  &  $ 3.0 \times 10^{-3}$ \\
\hline
\end{tabular}
\end{center}
\caption{Constraints on various $\varepsilon$'s from $\ell
\rightarrow \ell\ell\ell$, one-loop $\ell \rightarrow \ell \gamma$,
and $\mu^+ e^- \rightarrow \mu^- e^+$ processes. The experimental
bounds have been obtained from
Refs.~\cite{Willmann:1998gd,Amsler:2008zz}.} \vspace{-0.5cm}
\label{tab:constraints}
\end{table}
Notice that the stringent bounds listed in
Table~\ref{tab:constraints} are related with at least one
off-diagonal entry of $m_{\nu}$. In order to receive sizable NSIs
effects in neutrino experiments and avoid large LFV processes at the
same time, one expects the flavor non-diagonal parts of $m_\nu$ to
be relatively small. Hence, $m_\nu$ should take an approximately
diagonal form, which is quite favorable in the case of a nearly
degenerate (ND) neutrino mass spectrum $m_1 \simeq m_2 \simeq m_3$.
Then, $m_\nu$ approximates to a unit matrix, and the only relevant
NSI parameters are $\varepsilon^m_{ee}$ and
$\varepsilon^{e\mu}_{e\mu}=(\varepsilon^{\mu e}_{\mu e})^{*}$.
Focusing on the ND case, one can at leading order neglect the
neutrino mass-squared differences, so that the generic formula
(\ref{eq:epsilon}) is simplified to e.g.
\begin{eqnarray}\label{eq:approx}
\varepsilon^m_{ee}  & \simeq & -\frac{m^2_\Delta m^2_1}{8\sqrt{2}G_F
v^4 \lambda^2_\phi} \left| \sum^3_{i=1} U^2_{ei}  \right|^2 \ ,
\\
\label{eq:approx2} \varepsilon^{e\mu}_{e \mu}  & \simeq &
-\frac{m^2_\Delta m^2_1}{8\sqrt{2}G_F v^4 \lambda^2_\phi}
\left(\sum^3_{i=1} U^2_{ei} \right)^* \left( \sum^3_{i=1}U^2_{\mu i}
\right) \,.\;\;\;
\end{eqnarray}
where $U$ is the leptonic mixing matrix. The generic upper bounds on
the NSI parameters (\ref{eq:epsilon}) with respect to the lightest
neutrino mass $m_1$ are illustrated in Fig.~2, in which a normal
neutrino mass hierarchy $m_1<m_2<m_3$ is assumed. We take the
triplet Higgs mass $m_{\Delta} = 1 ~ {\rm TeV}$, which is a typical
value within the sensitivity range of the LHC. As for neutrino
mixing angles and mass-squared differences, we adopt the values from
a global fit given in Ref.~\cite{Schwetz:2008er}, and allow all the
CP violating phases to range from 0 to $2\pi$. Similar constraints
can be obtained in the case of an inverted neutrino mass hierarchy,
since the NSI parameters are only sensitive to $m_{1}$ in the ND region.

\begin{figure}[]
\begin{center}
\vspace{-0.1cm}
\ensuremath{\vcenter{\hbox{\includegraphics[width=8cm,height=5.5cm]{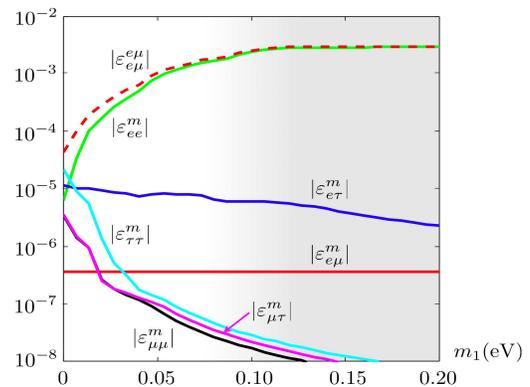}}}}
\vspace{-0.2cm}
\caption{The
upper bounds on the NSI parameters governed by Eq.~(\ref{eq:epsilon}) with respect to the lightest neutrino mass $m_1$.
Given $m_{1}$, the ratio of $m_{\Delta}/\lambda_{\phi}$ in
Eq.~(\ref{eq:epsilon}) (or $\lambda^{-1}_{\phi}$ if $m_{\Delta}$ is
fixed) is pushed up unless any of the bounds in Table
\ref{tab:constraints} is saturated.  The solid curves correspond to
constraints on $|\varepsilon^{m}_{\alpha\beta}|$ parameters, while
the dashed curve shows the upper bound on the
$|\varepsilon^{e\mu}_{e\mu}|$ (or $|\varepsilon^{\mu e}_{\mu e}|$)
parameter. Equations~(\ref{eq:approx}) and (\ref{eq:approx2}) yield
the asymptotic values of $|\varepsilon^{e\mu}_{e\mu}|$ and
$|\varepsilon^{m}_{ee}|$ in the ND region (shaded). The wiggles on
the curves are numerical artifacts given by the granularity of the
scan over the parameter space.} \vspace{-0.3cm} \label{fig:fig2}
\end{center}
\end{figure}

Clearly, only $\varepsilon^m_{ee}$ and $\varepsilon^{e\mu}_{e\mu}$
are significant at large $m_1$ regions, and they are similar in
size. The upper bound $|\varepsilon^{e\mu}_{e\mu}| \sim |\varepsilon^m_{ee}| \lesssim 3\times 10^{-3}$ 
can be obtained
according to  Fig.~\ref{fig:fig2}. As for the other NSI parameters,
the upper bounds are rather strong, which means that there is no
hope for them to be discovered in the near-future long baseline
experiments. Thus, $\varepsilon^{e\mu}_{e\mu}$ and $\varepsilon^m_{e
e}$ are the only relevant quantities to be taken into account in the
triplet seesaw model, and we will proceed to investigate their
effects at a neutrino factory and at the LHC.

\section{NSI effects at a neutrino factory and at the LHC}\label{Sec:3}

The concept of a neutrino factory is proposed to provide the
ultimate  high-precision measurement of the leptonic mixing
parameters and neutrino mass-squared differences. However, the pure and
intensive neutrino beams produced in muon decays make such a future
facility an ideal place to look for non-standard
physics.

\paragraph{Wrong sign muons at a near detector:}
The most striking signal of the underlying triplet model at a
neutrino factory corresponds to the processes
\begin{eqnarray}\label{eq:decay}
\mu^- \rightarrow e^- {\nu}_e  \overline{\nu_\mu} \ \quad {\rm and}\ \quad \mu^+ \rightarrow e^+  \overline{{\nu}_e}  \nu_\mu \ ,
\end{eqnarray}
leading to would-be observable rates of the ``wrong sign'' muon
tracks in a near detector \cite{Bueno:2000jy}, the so-called zero-distance effect
\cite{Langacker:1988up}. For instance, a significant admixture of
the first process in (\ref{eq:decay}) within the standard muon decay $\mu^-
\rightarrow e^-  {\nu}_\mu  \overline{\nu_e}$ would cause an observable
component of $\mu^{+}$ to appear along with the ``standard''
$\nu_{\mu}$-produced muons (in the $\mu^{-}$ run) well before the
oscillation effects $\overline{\nu_{e}}\to\overline{\nu_{\mu}}$ set in at longer
distances (of the order of a few hundred kilometers).
 Let us remark that the relative appearance rate of the ``wrong sign'' muons in a near
detector is approximately given  by
$|\varepsilon^{e\mu}_{e\mu}|^2$ for the $\mu^{-}$ run and by
$|\varepsilon^{\mu e}_{\mu e}|^2$ for the $\mu^{+}$ run,
respectively.
\begin{figure}[]
\begin{center}
\vspace{-0.5cm}
\ensuremath{\vcenter{\hbox{\includegraphics[width=7.5cm]{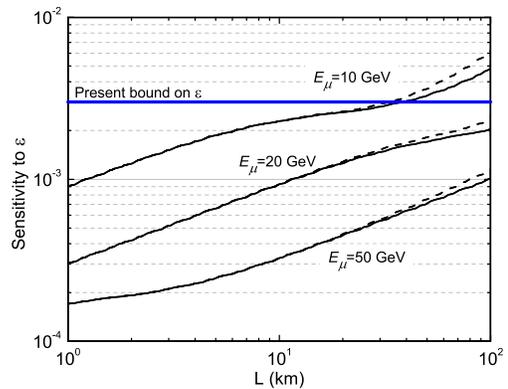}}}}
\caption{Sensitivity limits at 90~\% C.L. on the parameter $\varepsilon\equiv |\varepsilon_{e\mu}^{e\mu}|$ as a
function of the distance ($L$) from the source to the near detector.
The solid and dashed curves correspond to $\sin^2 2\theta_{13}=0.01$
and $0.1$, respectively.} \vspace{-0.6cm} \label{fig:fig3}
\end{center}
\end{figure}

In what follows, we illustrate the feasibility of observing such a
signal in the near detector of a typical neutrino factory setting.
For our numerical simulations, we use the GLoBES software
\cite{Huber:2004ka,Huber:2007ji} including a standard Abstract
Experiment Definition Language (AEDL) file to describe neutrino
factory experiments. We assume a setting with approximately $
10^{21}$ useful muon decays of each polarity and 4 years of
neutrino running plus other 4 years of antineutrino running. A
magnetized iron detector with perfect charge identification and a
fiducial mass of 1 kt are considered. We show in Fig.~\ref{fig:fig3}
the sensitivity to $\varepsilon\equiv|\varepsilon^{e\mu}_{e\mu}|$ as a
function of baseline length for a near detector. The parent muon
energies are labeled in the figure. It is obvious that a neutrino
factory provides an excellent sensitivity to probe this type of NSI
effects. The precision of $\varepsilon$ is limited by the baseline
(due to the oscillation effects), especially for a large
$\theta_{13}$. Thus, a distance $L \lesssim 10~ {\rm km}$ would be
favorable for the near detector.

Of course, our numerical demonstration in Fig.~\ref{fig:fig3} is
only an illustration, and in fact, a detector with fiducial mass 5 kt and $1+1$
years of running also result in similar sensitivities. For the sake of
completeness, a more systematic analysis of NSIs at a neutrino
factory is desirable. However, since the main purpose of this
letter is to clarify the situation about the potentially sizable NSI
effects in the triplet seesaw model, a detailed numerical study is
beyond the scope of this work and will be elaborated elsewhere.

\paragraph{Earth matter density profile uncertainties:}
In a long baseline experiment, the NSI parameter
$\varepsilon^m_{ee}$ enters the Hamiltonian as a shift in the Earth matter
density, which can be inferred in seismic measurements.
Therefore, precision measurements of $\varepsilon^m_{ee}$ are related to
how well the matter density uncertainty can be constrained. In
Ref.~\cite{Kopp:2008ds}, with the assumptions of the matter density
uncertainty around 1~\% and a sizable mixing angle $\sin^2 2\theta_{13} =
0.1$, a sensitivity of the order of a few percent can be achieved in a neutrino
factory with a two-detector setup. An improvement by a factor of
few would be possible if higher muon energies are considered. Once
an accurate geophysical estimate of the Earth matter density uncertainties becomes
available, one then may hope to perform a high sensitivity test on
$\varepsilon^m_{ee}$ in practice.

\paragraph{Like-sign di-lepton production at the LHC:}
It has been pointed out that, in the parameter region $ \lambda_\phi
v^2 / m^2_{\Delta} < 10^{-4} ~{\rm GeV}$, the dominant decay
channels of the doubly charged Higgs are $\Delta^{\pm\pm}\!\!\rightarrow\!
\ell^\pm_\alpha \ell^\pm_\beta$ (c.f. Ref.~\cite{Perez:2008ha}), and the partial
decay widths are governed by the corresponding triplet Yukawa
couplings $\Gamma(\Delta^{\pm\pm}\!\!\rightarrow\! \ell^\pm_\alpha
\ell^\pm_\beta)\propto \left|Y_{\alpha\beta}\right|^2 M_{\Delta} $.
According to the analysis above, these widths are
correlated to NSI parameters, since both these quantities are
sensitive to the neutrino mass spectrum. In the case of sizable
NSIs, $\Delta^{\pm\pm}$ produced at the LHC should predominantly decay into a pair of identical leptons. Similarly, one can expect
significant NSI effects if the branching ratios of decays to
identical leptons $\Delta^{\pm\pm}\!\!\rightarrow\! \ell^\pm_\alpha
\ell^\pm_\alpha$ would be  dramatically larger than for the other channels.

\section{Summary}\label{Sec:4}

We have argued that within the framework of a triplet seesaw model
sizable NSIs can naturally emerge as a consequence of a nearly
degenerate neutrino mass spectrum. Numerically, upper bounds for the effective couplings like
$|\varepsilon|\lesssim 3\times 10^{-3}$ have been obtained. 
We have studied in detail the potential of revealing some of these parameters at a near detector of a future neutrino factory and discussed in brief their 
possible effects on
the determination of the Earth matter density profile uncertainties and also collider signatures. 
We stress that NSI effects are generic features of the
seesaw models, and thus should be always properly dealth with. Combined
analysis of the electroweak precision tests, the 
neutrino oscillation experiments, and the future LHC results would be very helpful to figure out
the underlying physics behind the neutrino masses and mixing as well
as non-standard neutrino interactions.

\begin{acknowledgments}
We are grateful to Walter Winter and Toshihiko Ota for helpful discussions. The work was
supported by the Royal Swedish Academy of Sciences (KVA) [T.O.], the
G{\"o}ran Gustafsson Foundation [H.Z.], the Royal Institute of
Technology (KTH), contract no.~SII-56510 [M.M.], and the Swedish
Research Council (Vetenskapsr{\aa}det), contract no.~621-2005-3588 [T.O.].
\end{acknowledgments}


\end{document}